\documentstyle[epsfig,12pt]{article}
\newcommand{\beq}{\begin{equation}}
\newcommand{\eeq}{\end{equation}}
\newcommand{\bea}{\begin{eqnarray}}
\newcommand{\eea}{\end{eqnarray}}

\newcommand{\sm}{$SU(2)_L \times U(1)_Y $}
\begin{document}

\title{\vspace{-15mm}
       {\normalsize \hfill
       \begin{tabbing}
       \`\begin{tabular}{l}
  CERN--TH/95--344 \\
	 PM/95-25  \\
	 hep-ph/yymmmnn \\
	\end{tabular}
       \end{tabbing} }
       \vspace{8mm}
MSSM radiative contributions to the
WW$\gamma$ and WWZ form factors}
\author{A. Arhrib $^{a,b}$, J.-L. Kneur $^c$\thanks{On leave
from U.R.A 768 du C.N.R.S., F34095 Montpellier Cedex France.}
{}~~ and G. Moultaka $^a$}
\maketitle
\begin{center}
$^a$ Physique Math\'ematique et Th\'eorique\\
Unit\'{e} Associ\'{e}e au CNRS n$^o$ 040768,\\
Universit\'e Montpellier 2, F34095 Montpellier Cedex 5, France
\\
\vspace{5mm}
$^b$ Facult\'e des Sciences Semlalia, L.P.T.N., \\
B.P. S15, Marrakesh, Morocco\\
\vspace{5mm}
$^c$ CERN, Theory Division, \\
CH1211 Gen\`eve 23, Switzerland \\
\end{center}
\begin{abstract}
We evaluate one-loop contributions to the C and P conserving
$WW\gamma, WWZ$ form factors in the
Minimal Supersymmetric Standard Model (MSSM), and in
a more constrained Supergravity Grand Unified Theory
(SUGRA-GUT).
A systematic search of maximal effects in the available parameter
space, shows that at LEP2 energy MSSM contributions
can hardly reach the border of the most optimistic
accuracy expected on those
couplings, even for particles close to their production
thresholds.
At NLC energies, the effects are  more comfortably of the order
of the expected sensitivity, and may
therefore provide useful information on
MSSM parameter values which will not be available from
direct particle production. We also discuss briefly some variance with
other studies.
\end{abstract}
\vfill
\section{Introduction}
The $WW\gamma$, $WWZ$  Triple Gauge Couplings (TGC) will
be directly measured with a
decent accuracy of O(0.1) or better
at LEP2 \cite{bkrs1}
and, in a more remote future,
with an accuracy of O(10$^{-3}$) at the Next Linear Collider
(NLC) \cite{bkrs2},
i.e of the typical size of
electroweak radiative corrections in the latter case.
The effective Lagrangian parameterizing the most general
trilinear WWV interaction obeying C and P symmetries
is given by~\cite{gounaris} [V$\equiv \gamma$ or Z ]
\beq
\label{LTGC}
\L =  -ig_{VWW}[g_1^V\,V_\mu(W^{-\mu\nu}W^+_\nu -W^{+\mu\nu}W^-_\nu)
+\kappa_V V_{\mu\nu}W^{+\mu}W^{-\nu} \
 + {\lambda_V\over M^2_W}V^{\mu\nu}
W_\nu^{+\alpha}W_{\alpha\mu}^-]  \ ,
\eeq
where $g_{\gamma WW} = e$, $g_{ZWW} =e \cot\theta_W$, and
$g^V_1$, $\kappa_V$ and $\lambda_V$ are arbitrary, while
the SM \sm ~ gauge symmetry implies
$
g^\gamma_1  = g^Z_1 = 1$, $\kappa_\gamma =
\kappa_Z =1$, $ \lambda_\gamma = \lambda_Z = 0$,
at tree-level\footnote{
For $q^2 \neq 0$ (as it is in fact
relevant here), the arbitrary coefficients in
eq.(\ref{LTGC}) should be understood with a form-factor dependence, i.e
$g^V_1(q^2)$, $\kappa_V(q^2)$ and
$\lambda_V(q^2)$.}.

While such ``anomalous"
TGC are often purposed for parameterizing possible
{\it tree-level} deviations from the non-abelian Standard Model
(SM) gauge vertex, it is
worth to emphasize that any renormalizable extension of
the SM
(and indeed the SM itself), gives non-trivial
contributions to the TGC
at the radiative correction level\footnote{
Radiative corrections contribute to the
C, P and CP violating TGC as well. We
concentrate on C, P conserving contributions, since the
sensitivity to the C, P violating ('anapole') coupling is expected to be
less (by almost an order of magnitude)\cite{bkrs1,bkrs2}, and
the CP-violating TGC get radiative contributions only at the
two-loop level in the SM or MSSM
(provided that the soft susy terms are real).}.
But the generally expected decoupling \cite{appelquist,susydecouple}
of heavy new particles, plus the inherent
appearance of typical
$(4 \pi)^{-2} \simeq$ 6 10$^{-3}$ from loops,
lead to a largely consensual prejudice
that such radiative effects may be
generally small\cite{einhorn}, in particular most likely
below the reach of LEP2 measurements.
However the fact that
some of the supersymmetric partners could be relatively
light
give a complicated form factor dependence, threshold effects, etc...,
which may substancially enhance the overall
rough estimate above.
Given the plausibility of the MSSM as a New Physics candidate,
it is anyhow important to carry in some detail an exact
evaluation of such virtual contributions, ascertaining eventually their
irrelevance to LEP2 studies,
and examining in quantitative terms their more likely relevance at NLC.

There have been in fact numerous evaluations of virtual
contributions
to TGC in the past, both in the
SM~\cite{bgl,couture1,argyres,pinch2,fleischer}
and
supersymmetry\cite{bilchak,couture2,lahanas,alam}.
Most of these
calculations were however carried within some approximation (e.g
no $Q^2$-dependence\cite{bgl,couture1,bilchak,couture2},
massless fermions\cite{
bgl},
exact supersymmetry\cite{bilchak}, etc).
So far, the most complete analysis for the MSSM was performed
in refs.\cite{lahanas} and \cite{alam}.
In \cite{lahanas} the authors gave general analytic
expressions for vertex contributions, but considered only the much more
constrained SUGRA-GUT scenario in their numerical illustrations.
Moreover, most of the previous
analyses neglected the
box contributions (apart from the ones which are crucial
to gauge-invariance issues, see section 3.2 below).
An exception is ref.~\cite{alam}, where
the {\it full} one-loop MSSM contributions
to the $e^+e^- \to W^+W^-$ process were evaluated.
Although these contributions implicitely contain TGC as a part,
and give definite quantitative informations on the size of full
MSSM corrections to that process, they are
not expressed in terms of the parameters in eq.(\ref{LTGC}),
which will be determined from the data in addition to the
measurement of the $e^+e^- \to W^+W^-$ cross-section.
It turns out to be
difficult to extract from those results
the parameters in (\ref{LTGC}) without redoing most
of the calculation. \\ The purpose of
this letter is thus twofold: First we extend the work of Lahanas
and Spanos (comparing by-the-way our results to theirs), by exploring
in addition the unconstrained MSSM parameter space, in order to
look for possible experimentally measurable effects at LEP2 and NLC.
Secondly, we will illustrate
with one partial but
unambiguous (i.e gauge-invariant)
representative case what TGC contributions
can be expected from boxes.
This raises in fact some
general
questions on the issues of both a gauge-invariant and unique
definition of such TGC form factors.
%
%
\section{Survey of relevant ingredients of the MSSM}
As is well-known, the MSSM Lagrangian (restricted here to the R-parity
conserving case)
can be written as a supersymmetric part plus a (soft) supersymmetry
breaking part, ${\cal L}_{MSSM} = {\cal L}_{susy} + {\cal L}_{soft}
$.
${\cal L}_{susy}$ involves the  $SU(3) \times SU(2)_L \times U(1)_Y$
vector supermultiplets (gauge-bosons and their gaugino partners)
and chiral supermultiplets (Higgs scalars and their Higgsino partners,
leptons (quarks) and their slepton (squark) partners).
The supersymmetry-breaking part
${\cal L}_{soft}$ involves couplings among the scalars as well as the
phenomenologically
necessary splitting within each supermultiplet. For details
we refer to \cite{mssm,sugragut}. We simply list here the set of free
MSSM parameters that
we found convenient to choose
in our subsequent analysis: \\
$\bullet$
$\tan\beta \equiv v_u/v_d $, the ratio of the two Higgs--doublet
vacuum expectation values; \\
$\bullet$ the charged Higgs mass, $M_{H^+}$, which together with $\tan\beta $
determines (at tree-level) the CP-odd scalar
mass $M_A$,
the CP-even scalar masses $M_{h,H}$
and the mixing angle
$\alpha$ defining physical scalar states
(whereas the heavy top mass dominantly contributes to the
radiative corrections which largely
modify those tree-level mass values\cite{zwirner});
\\
$\bullet$ the $H_d$-$H_u$ mixing
parameter $\mu$, appearing in the MSSM
scalar potential (and entering also
the gaugino mass matrices); \\
$\bullet$ the soft gaugino mass terms $M_1$, $M_2$\footnote{
note that the gluino mass term, $M_3$, does not contribute to the
TGC at the one-loop level}, which together with $\mu$
and $\tan\beta$ determine the chargino and neutralino mass eigenstates
and couplings to the gauge bosons;
\\
$\bullet$ finally all the soft squark and slepton mass terms,
which due to the mixing between the left and right sfermions involve
two mass eigenstates and a corresponding mixing angle:
$\tilde m^i_1$, $\tilde m^i_2$, $\tilde \theta^i$
for any different squark and slepton flavor $i$.
\\
The unconstrained MSSM clearly gives a huge number of parameters
to consider if no
further theoretical assumptions are made. One attractive scenario
is thus to consider the MSSM
as emerging from a SUGRA-GUT\cite{sugragut}:
in this case \footnote{we disregard here the possibility
of non-universal soft terms}
one has, at the GUT scale, a universal scalar mass scale,
$m_0(\Lambda_{GUT})$
for all sfermion mass terms; a universal gaugino
mass, $M_1= M_2 = M_3 = m_{1/2}(\Lambda_{GUT})$, and a unique
trilinear soft
term, $A_0(\Lambda_{GUT})$ (the latter only
enters the sfermion mass mixing
terms as far as TGC contributions are concerned).
The
various soft terms for any flavor at a chosen scale
are then determined by the Renormalization Group
(RG) running. An additional attractive feature is
the possibility of radiative breaking of $SU(2)_L \times U(1)$
\cite{sugragut} within this scenario, which
we will take into account when considering SUGRA-GUT
contributions in our numerical illustrations
\footnote{to determine the spectrum from RG running,
we use for definiteness the procedure given in
ref. \cite{barger}}.
The remaining parameters accordingly
are $\tan \beta $, the top mass (which we
fix however to $m_{top} =$ 175 GeV in the following),
and the sign of $\mu$.
It would be of course interesting if SUGRA-GUT gave
a distinct signature with respect to the unconstrained MSSM.
In section \ref{numerics}
we illustrate the behavior of TGC as a function of the various
MSSM or SUGRA-GUT parameters listed above.

\section{Extracting TGC contributions from loops}
In momentum space the vertex issued from the effective Lagrangian in
(\ref{LTGC}) reads
\bea
\label{vertex}
\Gamma^V_{\mu \alpha \beta} = i g_{VWW} \{f_V
[2g_{\alpha\beta} \Delta_\mu +4(g_{\alpha\mu} Q_\beta-g_{\beta\mu}
Q_\alpha )] \\ \nonumber
+2\Delta\kappa^\prime_V(g_{\alpha\mu}Q_\beta -g_{\beta\mu}Q_\alpha)
+4{\Delta Q_V \over M^2_W} \Delta_\mu(Q_\alpha Q_\beta -
g_{\alpha \beta}{Q^2\over 2}) \},
\eea
where
$2Q_\mu$, $(\Delta-Q)_\alpha$, and $-(\Delta+Q)_\beta$ designate the
four-momenta and Lorentz
indices of the {\it incoming} $\gamma$ (or $Z$),  $W^+$, and
$W^-$, respectively.

The coefficients in \ref{vertex} are related
to the original TGC parameters in (\ref{LTGC}) according to
\bea
\label{tgclink}
\Delta\kappa^\prime_V \equiv \kappa_V -1 +\lambda_V \;=\Delta \kappa_V
+\lambda_V; ~~~~~
\Delta Q_V \equiv -2 \lambda_V.
\eea
Though trivial, the relations in (\ref{tgclink}) are important to
remember when comparing the radiative contributions from a given model,
generally more conveniently evaluated in terms of $\Delta\kappa^\prime_V$,
$\Delta Q_V$ \cite{bgl}--\cite{lahanas}, with the constraints obtained from
simulated
data, more traditionally given as bounds on $\Delta\kappa_V$
and $\lambda_V$.
Note however that we disagree with \cite{lahanas,argyres,pinch2} on an overall
minus sign difference in
(\ref{vertex}) (thus in $\Delta \kappa_V$, $\lambda_V$).
Our definitions in (\ref{vertex}), (\ref{tgclink})
are consistent with SM tree-level couplings and, in particular,
 with the parametrization in
\cite{bkrs1}-\cite{gounaris} and \cite{bgl}.
\subsection{Naive vertex contributions and gauge invariance}\label{naive}
To extract
from any triangle graph the contributions to $\Delta\kappa^\prime_V$, $\Delta
Q_V$ in eq.(\ref{vertex}), we adopt a systematic procedure to deal with
the large number of Feynman graphs contributing in the MSSM, avoiding as much
as possible manipulation by hand. The relevant graphs
are first evaluated
analytically, using FeynArts and FeynCalc packages
\cite{feyncalc} including a full MSSM Feynman rules code \cite{abdes}.
Contributions to (\ref{vertex}) are then
systematically extracted by algebraic manipulation with the help of
Mathematica\cite{mathematica}. We then can proceed to
a purely numerical
evaluation in terms of the standard
Passarino-Veltman functions\cite{pave}, with the help of
FF-package \cite{oldenborgh}. We keep as well
intermediate analytical
expressions in terms of integrals over two Feynman parameters $x$, $y$,
which turn out to be very compact and thus convenient to compare with
similar analytical expressions previously obtained in the literature
\cite{bgl, argyres, lahanas}. At this stage we obtain
a complete agreement with
the analytic results given in ref. \cite{lahanas} for the MSSM triangle
graph contributions, apart from an overall minus sign as mentioned above.
As noticed by these authors, the contributions from ordinary fermions
differ however from previous results in the literature.
In addition, several consistency cross-checks of our results were done,
like e.g
the vanishing of the total contributions to $\Delta Q_V$ for
{\it exact} supersymmetry (for arbitrary $Q^2$),
the decoupling behavior, $\Delta \kappa^\prime_V$, $\lambda_V$ $\to 0$
 for large mass values (in the limit of MSSM parameters where it is
expected to hold~\cite{susydecouple}), etc.

A problem which one immediately encounters is that
the vertex graphs with virtual gauge bosons
depend on the gauge
fixing parameter, $\xi$ in R-$\xi$ gauges.
These vertices need to be combined with parts
of boxes and self-energies to become gauge-invariant.
A general and non-ambiguous way of making such
a gauge-invariant separation would be to fully project the on-shell amplitude
on a complete operator basis
(see for instance \cite{denner}), which would define
by the same token the various WWV form factors (plus some remant, non-TGC
contributions~\cite{fleischer}).
Alternatively it was proposed
to extract the desired gauge-invariant contributions
directly, by so to speak `pinching' the irrelevant propagator lines
\cite{pinch1}. When applied to the TGC this
allows to calculate only vertex-like, three-point functions,
and was shown\cite{pinch2}
to lead to
a number of well-behaving features and properties expected from radiative
corrections
(simple Ward identities, good unitarity behavior, infra-red finiteness etc).
Accordingly in our calculation we have included the pinch parts
of box counterparts of the gauge-dependent vertices, and
verified the aforementioned properties. \\
Now despite its simplicity
and efficiency, the pinch technique raises some questions about the
definition, universality, and extraction procedure of TGC quantities:
by construction additional gauge invariant box contributions to TGC
are left over. We shortly address this issue in section (\ref{non-pinch})
(a detailed treatment
will be given elsewhere~\cite{inpreparation}).
\section{TGC contributions in the MSSM}\label{numerics}
With the latter cautionary remarks in mind, we proceed
to the numerical illustrations of the TGC from vertices plus the
pinched box parts forming a gauge-invariant combination.
We restrict here the study to $\Delta
\kappa^\prime_{\gamma ,Z}$ and $\Delta Q_{\gamma ,Z}$
among other anomalous couplings, since these (together with $g^Z_1$ in
eq. (\ref{LTGC})) are expected to be measured with
the best accuracy at LEP2 and NLC~\cite{bkrs2}.
To illustrate the sensitivity to the
various parameters in the unconstrained MSSM case, we give
separately contributions
from the Higgses (fig. 1), sfermions (fig. 2),
and gauginos (fig. 3)~\footnote{Note that those
three different sources of TGC contributions
do not mix at the one-loop level.},
as functions
of the parameters that we found the most illustrative in each case (see
figure captions for details).
A few additional comments may be useful: \\
In fig.1, the Higgses contribution becomes practically constant
for $M_{H+} >$ 200 GeV and/or $\tan\beta >$ 6--8, approximately:
for those values of $\tan\beta$,
$M_h \to M_Z$ (+rad. corr.) and $M_H \simeq const. M_{H+}$, so
there practically only remains the contribution from the approximately
constant,
light Higgs mass, $m_h$,
while the other Higgses give decoupling contributions for large $M_{H+}$.\\
In fig.2, the sfermion contributions are shown. There
are in principle so many arbitrary sfermion masses in the unconstrained MSSM
that we have to
make some choice in order to illustrate sfermion mass dependence.
Accordingly, guided by the mass values obtained when searching
for maximal effects (see the discussion below),
we show here the variation
of the {\it total} sfermion contributions versus one of the stop mass
eigenvalues,
$\tilde{m}_{t1}$, with other squark and slepton masses related to
$\tilde{m}_{t1}$
in a definite way (see figure caption for details).
Of course we have tried many other configurations, and
in particular since sfermion contributions can be either positive or
negative, depending on the squark/slepton charges, one can obtain for
the total sfermion contribution
almost any possible value between the maximal and minimal ones, respectively
given in table 1. As a general behavior however, we mention that the
dependence upon the mixing angle is quite mild (with a maximum in magnitude
for zero mixing); also the effects increase for increasing mass splitting
between any up and down components (with positive effects dominated by the
slepton contributions and negative effects dominated by squark contributions).
\\
The gauginos contributions are illustrated in fig. 3 as function of the
parameter $\mu$, for some representative
choice of the other relevant parameters
\footnote{Actually we should exclude
on figure 3 a central band in $\mu$
corresponding to the present (LEP) and future
(NLC) direct constraints on chargino/neutralino masses. We nevertheless kept
the effects
inside those bands for illustration.}
(see figure caption for details).
The maximal effects in $|\Delta \kappa_{\gamma, Z}|$ are
always due to chargino or neutralino threshold effects,
and in some cases even anomalous threshold effects show up,
as we checked explicitely (one example of the latter
corresponds to the small discontinuities in $\Delta \kappa'_Z$ case b) of
figure 3.B.).
Note also that at LEP2 (Fig.3.A),
$\Delta Q_{\gamma, Z}$ can become comparable to $\Delta \kappa_{\gamma, Z}$ for
large
$\mu$ and away from threshold effects.
\\
We have also performed a
rather systematic search of maximal effects, with the help of
standard minimisation tools~\cite{minui}.
Some typical 'large' effects are illustrated in table 1, both
for separate sectors and total contributions.
SM contributions are also given for comparison in table 1,
for $m_{top} =$ 175 GeV and $m_{Higgs}=$ 0.06 --0.6 TeV.\break
One should note that the maximal effects in the MSSM are mostly (though not
entirely) due to threshold effects, corresponding to the
very unlikely case where most particle masses are very close to their direct
production thresholds.
For sfermion contributions however, as mentioned above
large mass splittings between up and down
components of a same doublet substancially increase the
contributions. This can be understood by noting that no particular
decoupling
property is expected in that case: actually those contributions tend
to a constant for very large mass splitting between up and down components.
Furthermore, the extremal values illustrated in
the figures, both for positive or negative contributions, are quite
close to what we obtained from maximization.
In table 1 we considered maximal effects only at $\sqrt s =$ 190 GeV.
At 500 GeV  it is less compelling to look for maximal effects,
given the trend of the contributions in this case.
We simply quote here
the (approximate) extremal values, in units of $(g^2/16\pi^2))$,
for the {\em total} contributions obtained at this energy:
\beq
\Delta\kappa^\prime _{\gamma} =-1.955 ;\;\;\; \Delta\kappa^\prime_{Z} =-0.99
\; . \eeq
In all those illustrations we took into account as much as possible
already existing constraints on some parameters, like the lightest Higgs
mass, sfermions, and gauginos lower bounds,$\tan\beta$ constraints
etc.
\\
As a general remark, the magnitude of
the effects tends to decrease at 500 GeV, with respect to LEP2 energies:
this is indeed expected, once thresholds are crossed, in accord with
the good unitarity
behavior expected from a renormalizable gauge theory.  Fortunately the
expected accuracy of TGC measurements greatly increases at
NLC~\cite{bkrs2},
which more than compensates the latter effect. \\
In addition,
we show in fig. 4 the {\it total} contributions for one particular
SUGRA-GUT choice of parameters, the
no-scale scenario~\cite{noscale}, which was chosen to illustrate one
constrained example in contrast with the general cases above.
We have compared our results
with the ones from ~\cite{lahanas} for different SUGRA-GUT
parameter choices.
Apart from the already mentioned overall sign difference everywhere,
we have some discrepancies, namely for $\Delta \kappa^\prime_\gamma $
and $\Delta Q_Z$, which are very pronounced at high energy (500 GeV).
On the other hand we get very good agreement (given the completely
different numerical tools used, and, more essentially, the slightly
different way of evolving the SUGRA-GUT parameters from $\Lambda_{GUT}$ to
low energy scales) with their results
for $\Delta\kappa^\prime_Z$ and $\Delta Q_\gamma$.
Accordingly, as far as we can see, the discrepancies cannot be traced to the
slightly different
way of evolving the parameters with the renormalisation group in our
analysis. Our results
show a more rapid decoupling behavior at high energy,
though that does not by itself guarantee correctness. \\
In summary,
one can hardly expect to see any MSSM TGC at LEP2, where
even the most optimistic accuracy expected, $\vert \Delta \kappa_\gamma
\vert < 0.02 \simeq $ 6-7 $(g^2/16\pi^2)$~\cite{bkrs1},
hardly compares with the maximal effects\footnote{There are
larger radiative corrections to the $e^+e^- \to W^+W^-$ process, especially
at high energies, which are essentially due to QED Initial State
Radiations (ISR)~\cite{denner,beenakker}.
Though those would formally contribute to the
$g^{\gamma ,Z}_1$ TGC
in eq. (\ref{LTGC})~\cite{fleischer}, they should obviously not be taken
into account in our evaluation of New Physics virtual effects. In principle,
those large ISR effects can be corrected for before extracting TGC from data.}
in table 1.
In contrast the effects at NLC can be comfortably above
the expected accuracy for a reasonably large range of the parameter space
(taking $\vert \Delta \kappa_{\gamma , Z}
\vert < $ 10$^{-3}\simeq$ 0.3 $(g^2/(16\pi^2)$\cite{bkrs2}
as a reference accuracy at 500 GeV,).
In particular, even for the more constrained SUGRA-GUT scenario we obtain
effects above the accuracy limit, although the no-scale case illustrated
in Fig.4 does not give the largest possible contribution.
Accordingly, at a 500 GeV NLC it should even be possible to obtain
useful information on the MSSM parameter space in the range which will
not be accessible from direct production processes.
\section{Additional non-pinch box contributions}\label{non-pinch}
The previous picture is valid if the genuine (i.e non-pinch) box
contributions, generally omitted in most evaluation of TGC, are
truely negligible.
As mentioned in section \ref{vertex} above, by construction
the pinching takes from a box just what
is necessary to cancel the $\xi$-dependence of vertices, therefore
leaving out other possible gauge-invariant (box) contributions.
The resulting combinations of pinch boxes plus vertices give a TGC
contribution with s-dependence {\it only}~\cite{pinch2},
which in that sense is meant to be ``process-
independent" and universal.
However it was noted earlier in SM
\cite{fleischer},
that once projecting the $e^+e^- \to W^+W^-$ one-loop
corrections over the {\it complete}
operator basis,
the obtained TGC clearly exhibit both
$s$ {\it and} $t$ dependence.
One thing is that, to our knowledge, there is no proof that {\it no other}
possible universal contributions from boxes are left out by the pinch
technique.
Even if the latter statement could be proven, one problem would persist,
since experimentally there are a priori no planned procedure to
distinguish 'universal' from 'non-universal'
TGC. So far all analysis
have extracted expected TGC constraints, from fitting angular
distributions for simulated data to theoretical expressions
assuming {\it t-independent} TGC\footnote{To
distinguish a t-dependence, one would need typically
to allow in the fit the TGC to be
different for different scattering angles, which would most likely
considerably reduce the expected accuracy on such TGC.}.
Therefore, an
unambiguous procedure to test a {\it specific
model} prediction via
TGC measurements, is to evaluate the full contributions
to the definite process where TGC are extracted.
This is of course a much more involved
program, but to illustrate here simply what one should expect in general,
we have evaluated one first partial but unambiguous (gauge-invariant)
contribution, the sum of boxes with one
internal chargino (resp. neutralino),
two internal sneutrinos (resp. selectrons) and one internal selectron (resp.
 sneutrino) , which do contribute to $\Delta \kappa'_\gamma$
and $\Delta \kappa'_Z$ yet cannot be obtained from
the pinch technique.
The results are shown
on fig. 5 at $\sqrt s =$ 500 GeV.
The effects are clearly comparable to the vertex contributions,
even when most of the chargino/neutralino masses are above threshold.
In contrast, this particular
box contribution is totally negligible at LEP2,
giving at most ${\cal O}(0.1 g^2/16\pi^2) \simeq$ 2.7 10$^{-4}$ TGC
at $\sqrt{s}= 190GeV $. Of course this is only
a partial contribution, so that no general conclusion can be drawn from it.
What may be interesting on that particular example is that
the $t$-dependence
is relatively smooth, so that neglecting t-dependence
in the fits may not introduce too much biases.
In any case, even if such issue is irrelevant for LEP2, given the
too small size of vertex contributions anyway, this example
should emphasize the need to evaluate
all other boxes at NLC energies: there,
any source of TGC contribution is likely to be sizeable and, accordingly,
one should be careful to sum all
the relevant contributions if one wants a precise
comparison to the data.The complete evaluation of boxes,
together with a more detailed illustration of both MSSM and SUGRA-GUT
contributions,
is at present under investigation~\cite{inpreparation}.

{\bf Acknowledgments} \\
A.~A. and G.~M. acknowledge partial support from a EC contract
CHRX-CT94-0579.
J.-L. K. acknowledges support from a
CERN fellowship.
We are thankful to M. Carena, F. Jegerlehner, J. Papavassiliou, F. M. Renard
and C. Wagner for
useful discussions.
We also thank A. Lahanas and V. Spanos for useful exchange of information.

 \vfill \eject

{\large \bf Table Caption}

{\bf Table 1}: Maximal contributions to $\Delta\kappa^\prime _V $ and
$\Delta\kappa_V (\equiv \Delta\kappa^\prime _V +\Delta Q_V/2)$
in unconstrained MSSM, at 190 GeV ({\em in units of} $(g^2/16\pi^2))$.
The correponding values of input parameters are indicated (all masses
are in GeV). $\tilde{U}$ and $\tilde{D}$ denote generically all up
and down squarks respectively.
The SM (total) contributions are also shown for two values of $M_h$.\\

{\large \bf Figure Captions}:

{\bf Fig.${\bf 1.A}$}: Total MSSM Higgs contribution at $\sqrt s = 190 GeV$
to
$\Delta\kappa^\prime_{\gamma ,Z}$ versus $M_{H+}$ for different values
of $\tan\beta$. In all the plots the ordinate numbers are in units of
 $g^2/16\pi^2$;\\

{\bf Fig.${\bf 1.B}$}: same as for Fig. 1.A at $\sqrt s = 500 GeV$\\

{\bf Fig.${\bf 2.A}$}: Total squark and slepton contribution to
$\Delta
\kappa^\prime_{\gamma}$ and $\Delta \kappa^\prime_Z$
versus $m_{\tilde{t}_1}$, at
$\sqrt{s}= 190 GeV$  with the following mass spectrum configuration :
$m_{\tilde{t}_1}=m_{\tilde{t}_2}=m_{\tilde{U}_1}=m_{\tilde{U}_2}=m_{\tilde{{
\it l}}_1}=m_{\tilde{{\it l}}_2} $
and $m_{\tilde{t}_1} + m_{\tilde{\nu}_L}\simeq m_{\tilde{t}_1} + m_{\tilde{D
}_{1,2}}=1.09 TeV $ ;all left-right mixing angles are vanishing;\\

{\bf Fig.${\bf 2.B}$}: same as in Fig. 2.A at
$\sqrt{s}= 500 GeV$ except that now
$m_{\tilde{t}_1} + m_{\tilde{\nu}_L}= 1.245 TeV$ and $m_{\tilde{t}_1} + m_{
\tilde{D}_{1,2}}= 1.47TeV$.\\

{\bf Fig.${\bf 3.A}$}: Total chargino/neutralino contribution to  $\Delta
\kappa^\prime_{\gamma,Z}$ and $\Delta Q_{\gamma,Z}$
versus $\mu$, at $\sqrt{s}= 190 GeV$ with $M=100$ GeV, $M'=60$ GeV, $\tan(
\beta)=2$\\
%

{\bf Fig.${\bf 3.B}$}: Total chargino/neutralino contribution to  $\Delta
\kappa^\prime_{\gamma}$ and $\Delta \kappa^\prime_Z$
versus $\mu$, at $\sqrt{s}=500 GeV$, case a)  $M=190GeV$, $M'=70GeV$, $\tan(
\beta)= 2$,
                 case b)  $M=350GeV$, $M'=175GeV$, $\tan(\beta)= 2$;\\

{\bf Fig.${\bf 4.A}$}:$\Delta \kappa^\prime_{\gamma}$ and
$\Delta \kappa^\prime_Z $
at $\sqrt{s}= 190 GeV$, in no-scale SUGRA-GUT ($m_0=A_0=0$) as a function
of $m_{1/2}$. Both $\mu<0$ and $\mu>0$ cases are illustrated; \\

{\bf Fig.${\bf 4.B}$}:Same as in Fig.4.A for $\sqrt{s}= 500 GeV$ ;\\

{\bf Fig.${\bf 5}$}: An example of non-pinch box contributions to
$\Delta \kappa^\prime_{
\gamma}$ and $\Delta \kappa^\prime_Z$,
the sum of boxes with one
internal chargino (resp. neutralino),
two internal sneutrinos (resp. selectrons) and one internal selectron (resp.
 sneutrino) versus the $W^-$ production angle $\theta$, (defined
with respect to the beam axis)
in $e^+e^- \to W^+W^-$,
with $m_{\tilde{e}_1}=m_{\tilde{\nu}_L}=260 GeV$, zero left-right
mixing angle;
case a) $M=\mu=150 GeV$, $M'=100 GeV$, $\tan(\beta)=15$;
case b)  $M=M'=\mu= 250 GeV$, $\tan(\beta)=2$.

\vfill \eject

%
\begin{center}
{\bf Table 1}
\end{center}
{\small
\begin{table}
\begin{center}
\begin{tabular}{||l||c|c|c|c||}\hline \hline
 Contribution ($\sqrt s =$ 190 GeV) &
$\Delta\kappa^\prime _\gamma  $ & $\Delta\kappa_\gamma $ &
$\Delta\kappa^\prime _Z $ &$ \Delta\kappa_Z$ \\ \hline
$W,Z,\gamma $ +fermions ($m_t =$175) & $2.59$ & $2.338$ & $1.37$ & $1.13$ \\
Higgses ($\tan\beta =1.5$,$M_{H+}=95$) & $0.369$ & $0.344$ & 0.457 &0.427 \\
sfermions & 3.730 & 2.919 &1.561  & 1.133 \\
($m_{\tilde e_{1,2}}= m_{\tilde \mu_{1,2}} = m_{\tilde \tau_1}$
&  & & & \\
$= m_{\tilde U_1}
\simeq 92$ ; $ m_{\tilde D_1} = m_{\tilde \nu} \simeq 45$); & & & &  \\
gauginos & 0.750 & 0.889
&0.304 & 0.429  \\
($M \simeq 73, M^\prime \simeq 10, \mu \simeq -88$) &  & & & \\

Total MSSM & 7.439 & 6.490 &3.692  & 3.119  \\ \hline
SM ($m_t =$175, $M_h =$65--600) &$1.800$--$2.291$
&$1.530$--$2.039$ & $1.499$--$1.406$ &
$1.231$--$1.166$ \\

\hline \hline
\end{tabular}
\end{center}
\end{table}
}

\vfill \eject
\newpage

\begin{minipage}[t]{16.5cm}
\setlength{\unitlength}{1.in}
\begin{picture}(1,1)(0,9.)
\centerline{\epsffile{MHpfig-190-500.ps}}
\end{picture}
\end{minipage}
\vfill \eject
\begin{minipage}[t]{16.5cm}
\setlength{\unitlength}{1.in}
\begin{picture}(1,1)(0,9.)
\centerline{\epsffile{SFERMIONfig-190-500.ps}}
\end{picture}
\end{minipage}
\vfill \eject
\begin{minipage}[t]{16.5cm}
\setlength{\unitlength}{1.in}
\begin{picture}(1,1)(0,9.)
\centerline{\epsffile{CNfig-190-500.ps}}
\end{picture}
\end{minipage}
\vfill \eject
\begin{minipage}[t]{16.5cm}
\setlength{\unitlength}{1.in}
\begin{picture}(1,1)(0,9.)
\centerline{\epsffile{Noscalefig-190-500.ps}}
\end{picture}
\end{minipage}

\vfill \eject
\begin{minipage}[t]{16.5cm}
\setlength{\unitlength}{1.in}
\begin{picture}(1,1)(0,9.)
\centerline{\epsffile{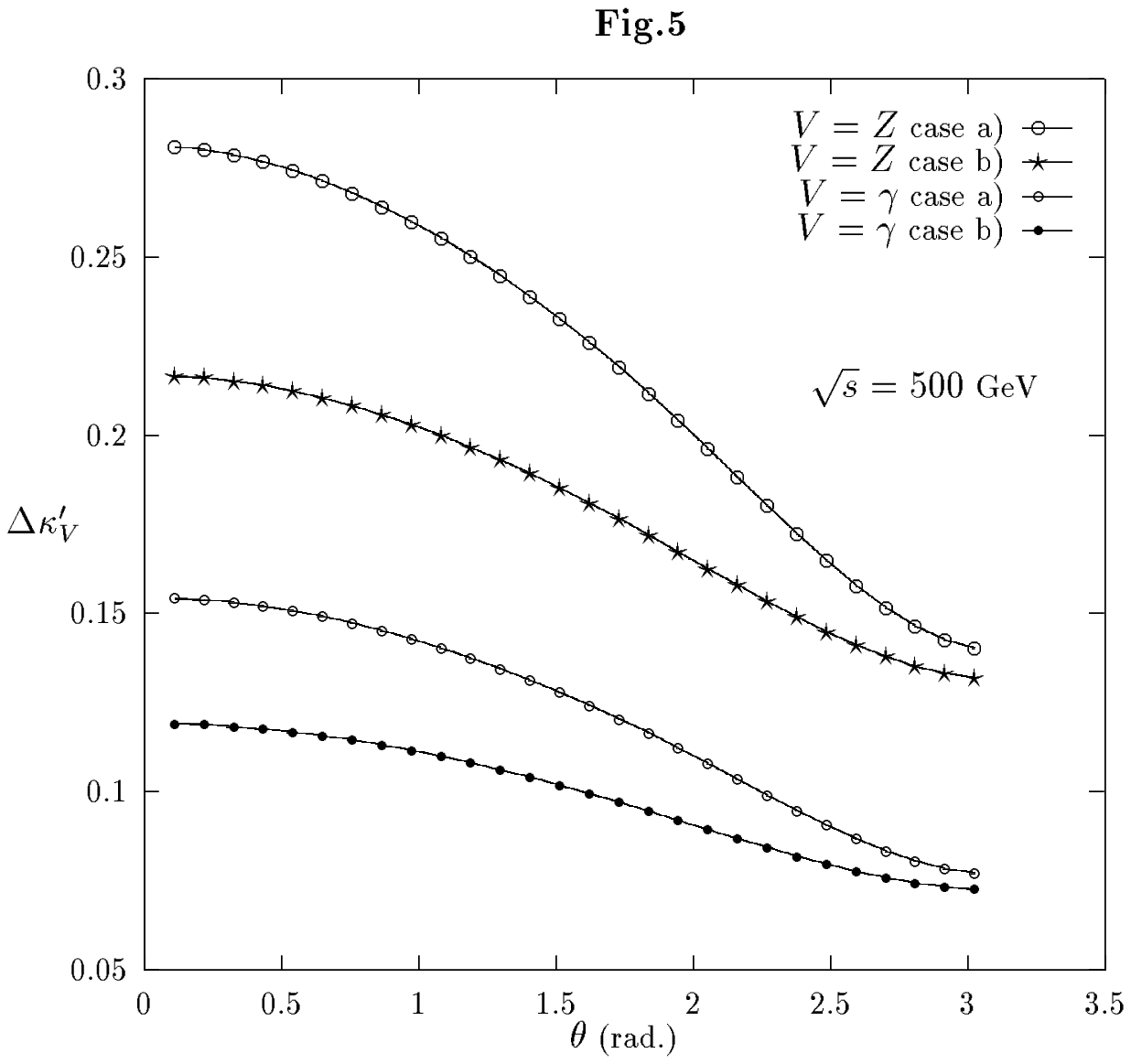}}
\end{picture}
\end{minipage}

\end{document}